\documentclass[aps,prl,showpacs,twocolumn]{revtex4}

\usepackage{here}

\usepackage{graphicx}
\usepackage{wrapfig}

\newcommand\pictc[5]{\begin{figure}
                   \centerline{
                   \includegraphics[width=#1\columnwidth]{#3}}
               \protect\caption{\protect\label{fig:#4} #5}
                \end{figure}            }

\newcommand\pict[4][1]{\pictc{#1}{!tb}{#2}{#3}{#4}}
\newcommand\rpict[1]{\ref{fig:#1}}

\newcommand\leqt[1]{\protect\label{eq:#1}}
\newcommand\reqtn[1]{\ref{eq:#1}}
\newcommand\reqt[1]{(\reqtn{#1})}

\newcounter{Fig}

\begin{document}
\begin{sloppy}

\title{Enhanced soliton transport in quasi-periodic lattices
with short-range aperiodicity}

\author{Andrey A. Sukhorukov}

\affiliation{Nonlinear Physics Centre and Centre for Ultra-high bandwidth Devices for Optical Systems (CUDOS),
Research School of Physical Sciences and Engineering,
Australian National University,
Canberra, ACT 0200, Australia}

\begin{abstract}
We study linear transmission and nonlinear soliton transport through quasi-periodic structures, which profiles are described by multiple modulation frequencies. We show that resonant scattering at mixed-frequency resonances limits transmission efficiency of localized wave packets, leading to radiation and possible trapping of solitons. We obtain an explicit analytical expression for optimal quasi-periodic lattice profiles, where additional aperiodic modulations suppress mixed-frequency resonances, resulting in dramatic enhancement of soliton mobility. Our results can be applied to the design of photonic waveguide structures, and arrays of magnetic micro-traps for atomic Bose-Einstein condensates.
\end{abstract}


\maketitle

Periodic structures can provide an efficient control of wave transmission and localization, making it possible to realize new physical regimes that are not allowed in homogeneous systems. Photonic crystals and Bragg gratings are used to tailor dispersion, diffraction, and emission of electromagnetic waves~\cite{Joannopoulos:1995:PhotonicCrystals, Russell:1995-585:ConfinedElectrons} and similar concepts have been developed and demonstrated for management of atomic Bose-Einstein condensates in periodic potentials~\cite{Cristiani:2002-63612:PRA, Anker:2005-20403:PRL}. 
Quasi-periodic photonic structures such as optical super-lattices and 2D quasi-crystals can offer even more flexibility in designing the properties of optical~\cite{Zoorob:2000-740:NAT} and atomic matter waves~\cite{Sanchez-Palencia:cond-mat/0502529:ARXIV, Scarola:cond-mat/0506415:ARXIV}. The modulation of quasi-periodic structures is defined by multiple incommensurable spatial frequencies, resulting in the lack of translational symmetry which is a feature of conventional periodic structures. This gives rise to an important effect of stronger field confinement at certain locations due to position-dependent coupling between the lattice sites, which can be utilized for various applications including efficient lasing~\cite{Notomi:2004-123906:PRL}. 

Many novel phenomena in periodic structures originate due to the modification of the linear spectrum which is separated into a number of transmission bands and band-gaps, where waves experience total reflection~\cite{Russell:1995-585:ConfinedElectrons}. For a one-dimensional lattice with period $d$, the gaps are centered at the frequencies corresponding to Bragg resonance condition for the wavevector component across the lattice, $2 k_x = m K$, where $m=\pm1,\, \pm2,\, \ldots$ is the order of resonance and $K = 2 \pi / d$ is the wavenumber of lattice modulation. In a quasi-periodic structure, primary gaps are defined by several dominant frequencies $K_j$, however there also appear multiple resonances at mixed frequencies $K_j \pm K_l$, resulting in a complicated spectrum containing multiple mini-bands and mini-gaps~\cite{Hollingworth:2001-36611:PRE}. This can have a detrimental effect on propagation of wave packets through the lattice, which can be  inhibited~\cite{Sanchez-Palencia:cond-mat/0502529:ARXIV, Scarola:cond-mat/0506415:ARXIV} or accompanied by strong radiation~\cite{Sumetsky:2002-332:OE}. It has been predicted that nonlinear wave self-action, which is known to appear due to atom-atom scattering in BEC or light-matter interactions in optical crystals, can suppress Anderson localization and allow wave transmission through disordered systems~\cite{Gredeskul:1992-1:PRP}. We find that such nonlinear transmission may be also possible in quasi-periodic lattices in the form of gap solitons (see Refs.~\cite{deSterke:1994-203:ProgressOptics, Eggleton:1996-1627:PRL, Fleischer:2003-23902:PRL, Mandelik:2004-93904:PRL, Neshev:2004-83905:PRL, Eiermann:2004-230401:PRL}, and references therein), however they still suffer from radiation losses that increase for stronger localized wave packets which spectra overlap mixed-frequency resonances. Resolution of these limitations may open new possibilities for realizing efficient transport of optical and matter waves in quasi-periodic structures.

In this Letter, we suggest a novel approach to the design of quasi-periodic structures. We show that mixed-frequency resonances can be suppressed through additional aperiodic lattice modulations which inhibit coherent back-scattering at specific frequencies. Aperiodicity if often associated with disorder, leading to the appearance of critically and fully localized states or Anderson localization~\cite{Soukoulis:1999-255:WRM, Scarola:cond-mat/0506415:ARXIV}. However, we present an analytical procedure which can be used to define, in a systematic way, quasi-periodic structures with short-range {\em deterministic aperiodicity} possessing regular (absolutely continuous) band-gap spectrum, where all eigenmodes are either extended or exponentially decaying. 
These `optimal' structures allow for unrestricted motion across the lattice of solitons strongly localized in any of the primary gaps, in full analogy with conventional periodic structures. 

\pict{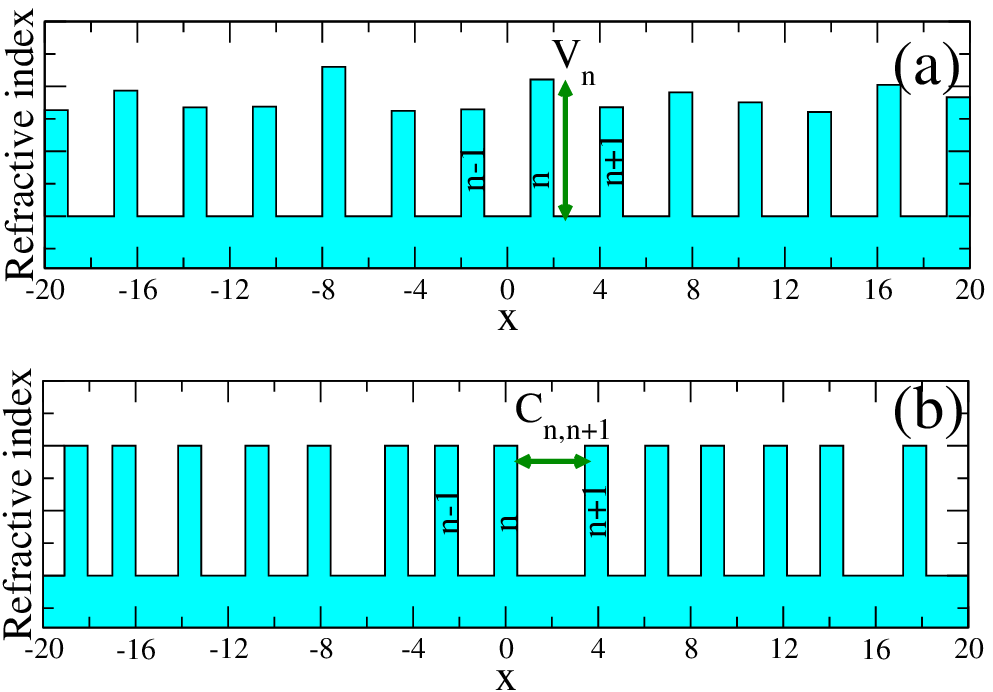}{sketch}{
Sketch of refractive index (or negative potential) in a lattice with a quasi-periodic modulation of
(a)~effective on-site potential, proportional to index contrast, and
(b)~coupling between the lattice sites, inversely proportional to waveguide separation.
}

We consider below wave transport and soliton motion in effectively discrete systems, such as arrays of optical waveguides or coupled-resonator structures in photonic crystals~\cite{Christodoulides:2003-817:NAT} or arrays of traps for matter waves~\cite{Trombettoni:2001-2353:PRL} where the wave transport is defined through tunneling between fundamental modes of the neighboring wells. 
This process can be described, in the tight-binding approximation, by the discrete nonlinear Schr\"odinger (or Gross-Pitaevskii) equation for the normalized amplitudes $E_n$,
\begin{equation} \leqt{dnls}
  \begin{array}{l} {\displaystyle 
    i \frac{\partial E_n}{\partial z} 
    + C_{n,n-1} E_{n-1} + C_{n,n+1} E_{n+1} 
  } \\*[9pt] {\displaystyle 
    + V_n E_n
    + {\cal F}( n, |E_n|^2 ) E_n
    = 0.
  } \end{array}
\end{equation}
Here $n$ is the number of waveguide or potential well, $z$ is the propagation distance or time, $C_{n m}$ are coupling coefficients, $V_n$ characterizes the detuning between the different sites, and ${\cal F}$ defines the nonlinear response [${\cal F}( n, 0 )\equiv 0$]. We note that the energy conservation law, $P = \sum_n |E_n|^2 = {\rm const}$ leads to the requirement that the coupling coefficients are symmetric, $C_{n, n-1} = C_{n-1, n}$. 

\pict{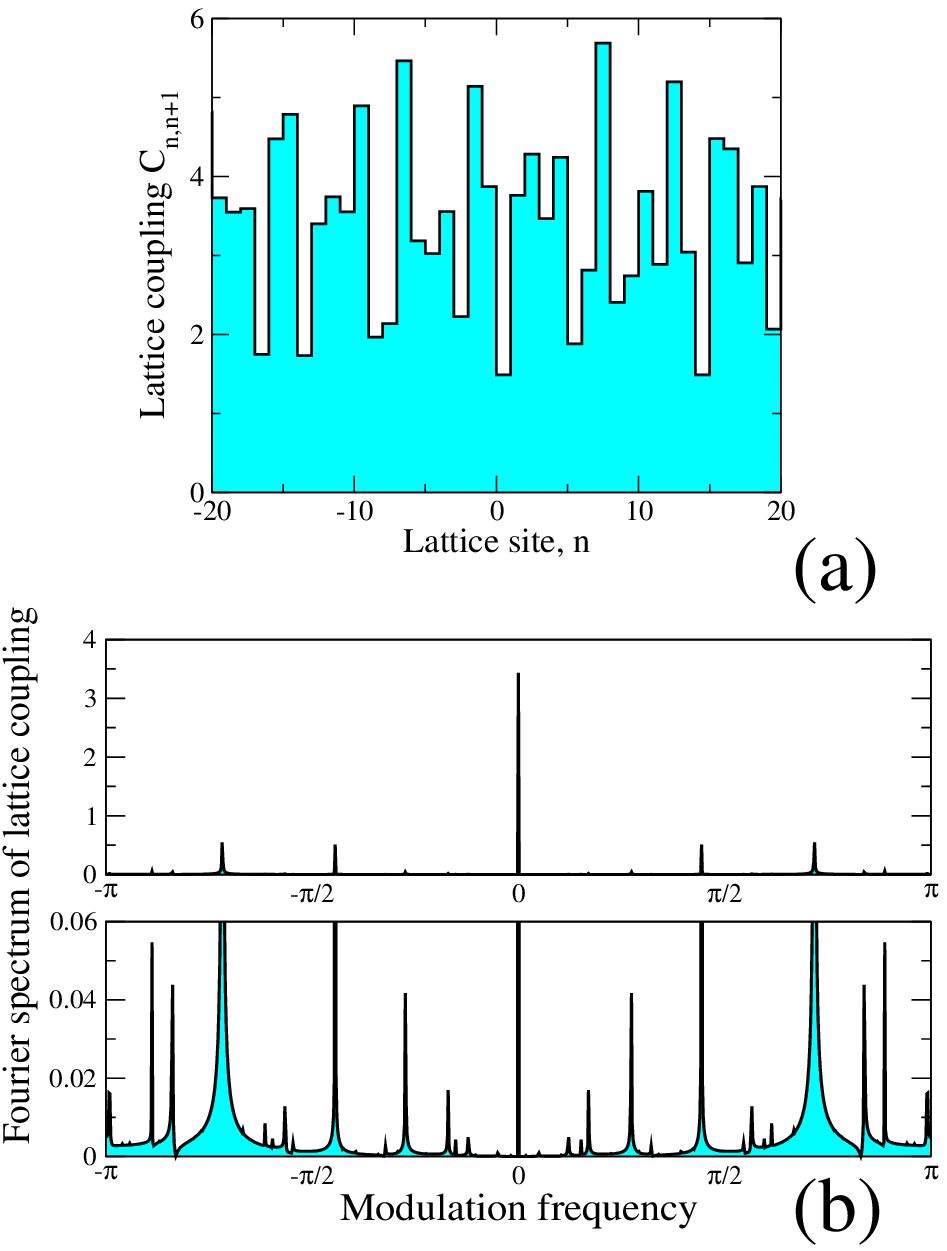}{coupling}{
Example of a lattice with quasi-periodic coupling and constant on-site potential [as sketched in Fig.~\rpict{sketch}(b)] calculated with Eq.~\reqt{coupling}:
(a)~Coupling vs. lattice site;
(b)~Fourier spectrum of the coupling coefficient, shown with different vertical scale in top and bottom plots.
}

The properties of discrete systems with a modulated potential $V_n$ but homogeneous coupling $C={\rm const}$ were extensively investigated with applications to the problem of electron transport and localization (see Refs.~\cite{Aubry:1980-133:RAR, Soukoulis:1982-1043:PRL, Sokoloff:1985-189:PRP, Lindquist:1994-9860:PRB, Soukoulis:1999-255:WRM}, and references therein). A special mathematical theory was developed to demonstrate that the linear band-gap spectrum remains regular (absolutely continuous) for harmonic modulations of $V_n$ below a certain threshold~\cite{Bourgain:2004-75:DCDS}, whereas aperiodicity usually leads to the appearance of dynamically or critically localized states which are neither extended nor exponentially decaying inside the structure, and this inhibits efficient propagation of wave packets apart from specific resonances. There is one well-known example of aperiodic structures defined through modified Thue-Morse sequences, which supports extended modes and fundamental lattice solitons, however their motion is also accompanied by radiation due to mixed-frequency resonances and gap solitons were not identified~\cite{Lindquist:1994-9860:PRB}. It remains an open problem whether there exist non-harmonic potentials which produce regular wave spectrum.

In engineered photonic structures such as waveguide arrays~\cite{Christodoulides:2003-817:NAT, Mandelik:2004-93904:PRL}, there is a flexibility in designing both the on-site potential, and also the coupling coefficients. These can be controlled independently by adjusting the waveguide width and their separation, as schematically illustrated in Figs.~\rpict{sketch}(a,b). Recently demonstrated arrays of magnetic micro-traps~\cite{Guenther:cond-mat/0504210:ARXIV} can be used to precisely design, in a similar way, the trapping potentials for atomic condensates. 
We show in the following that specially designed modulation of inter-site coupling can provide the optimal conditions for linear and nonlinear wave transport in quasi-periodic systems.

The model Eq.~\reqt{dnls} has an analog in the form of Ablowitz-Ladik system~\cite{Ablowitz:1999-287:PLA},
\begin{equation} \leqt{AL}
    i \frac{\partial {\bf \psi}_n}{\partial z} 
    + ({\bf \psi}_{n-1} + {\bf \psi}_{n+1}) ( C + |{\bf \psi}_n|^2 )
    = 0,
\end{equation}
where ${\bf \psi}_n = (\psi_n^{(1)}, \psi_n^{(2)}, \ldots)$ is a vector function defining the profiles of multiple interacting components. This is considered to be an integrable model~\cite{Ablowitz:1999-287:PLA}, where solitons would interact elastically (without radiation losses) with other nonlinear waves, including extended quasi-periodic modes. 
Although nonlinear modes induce asymmetric coupling in Eqs.~\reqt{AL}, and therefore do not correspond to real structures, we show that by identifying the effective modulations of lattice coupling induced by quasi-periodic nonlinear waves in the framework of Eq.~\reqt{AL}, we can then define equivalent `optimal' modulation for the original Eq.~\reqt{dnls}. Then, we consider solutions of Eq.~\reqt{AL} in the form of stationary modulated waves ${\bf \psi}_n(z) = {\bf \psi}_n(0) \exp(i \rho_n z)$, which profiles satisfy a set of difference equations,
\begin{equation} \leqt{ALst}
    - \rho_n {\bf \psi}_n 
    + ({\bf \psi}_{n-1} + {\bf \psi}_{n+1}) ( C + |{\bf \psi}_n|^2 )
    = 0.
\end{equation}
The waves with oscillating amplitudes ${\bf \psi}_n$ induce, in general, a quasi-periodic lattice with an effective modulation of coupling coefficients, $\widetilde{C}_{n, n\pm1} =  C + |{\bf \psi}_n|^2$. The lattice modulation can be controlled by choosing the number of modes, the values of $\rho_n$, and the initial conditions ${\bf \psi}_0$ and ${\bf \psi}_1$. Dynamics of small-amplitude excitations in such lattices are governed by a linear equation with the modulated coupling,
\begin{equation} \leqt{ALlin}
    i \frac{\partial \phi_n}{\partial z} 
    + (\phi_{n-1} + \phi_{n+1}) \widetilde{C}_{n,n\pm1}
    = 0.
\end{equation}
Most importantly, Eqs.~\reqt{ALst} are integrable~\cite{Suris:1994-281:PLA}, and therefore the band-gap spectrum of $\phi_n$ is completely regular, which is one of the key design requirements as outlined above. We now introduce a scaling transformation for the mode amplitude, $\phi_n \rightarrow \phi_n  \sqrt{\widetilde{C}_{n,n+1}}$, which preserves the regular band-gap spectrum and simultaneously transforms Eq.~\reqt{ALlin} into a linearized Eq.~\reqt{dnls} with symmetric coupling found as
\begin{equation}  \leqt{coupling}
  \begin{array}{l} {\displaystyle 
     C_{n, n+1} = C_{n+1, n} = \sqrt{ \widetilde{C}_{n,n+1} \widetilde{C}_{n+1,n} }
  } \\*[9pt] {\displaystyle 
         = \sqrt{ (C + |{\bf \psi}_n|^2) (C + |{\bf \psi}_{n+1}|^2) }.
  } \end{array}
\end{equation}
This is the key expression which defines an optimal coupling that can be implemented in real periodic structures.

\pict{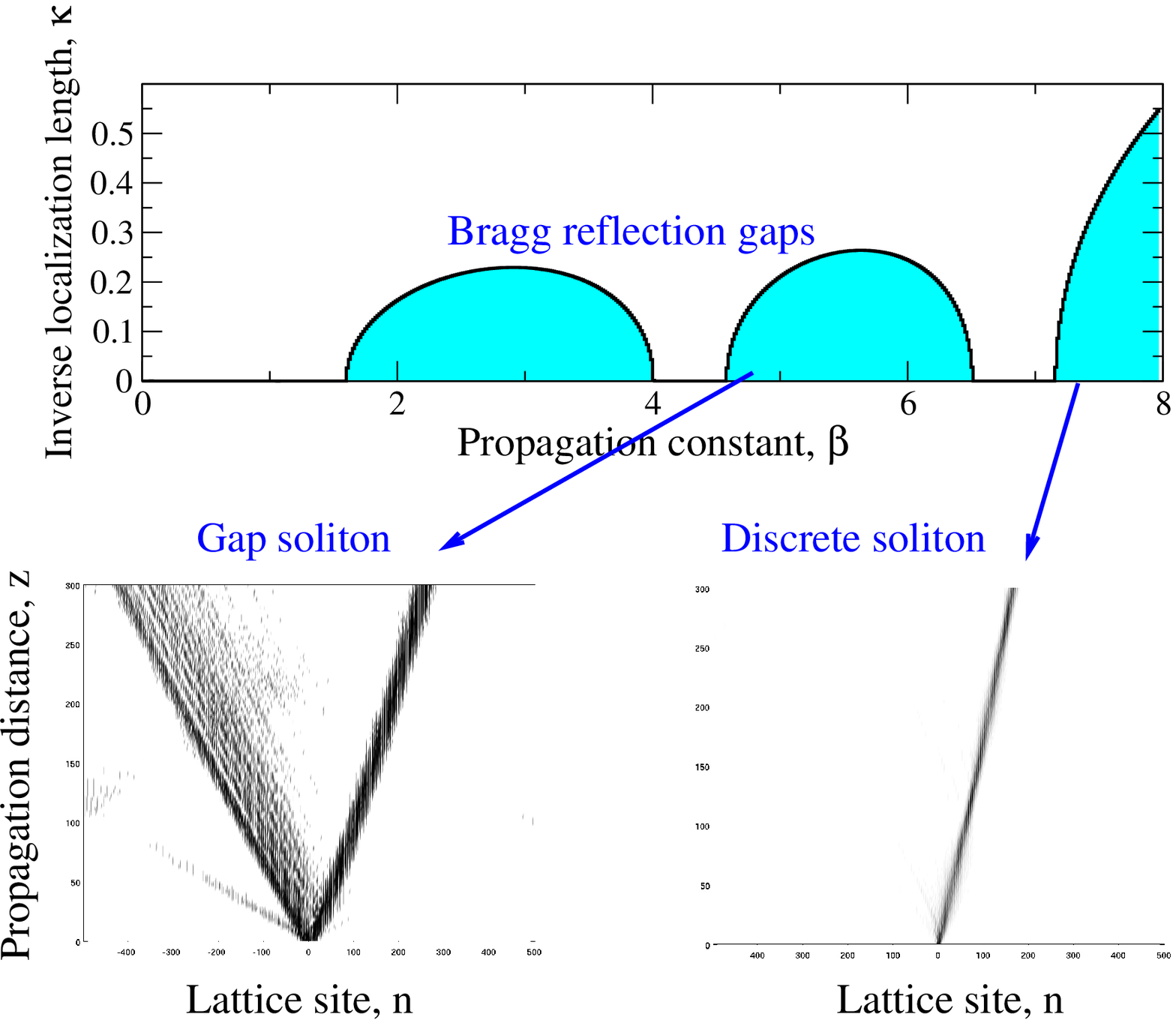}{soliton}{
Top: Band-gap diagram shown as inverse localization length vs. the propagation constant.
Bottom: Moving discrete and gap solitons corresponding to the total internal reflection and Bragg-reflection gaps as indicated by arrow. 
Solitons are excited by a single tilted beam (discrete) or two tilted beams at the Bragg angle. The lattice parameters correspond to Fig.~\rpict{coupling}.
}

\pict{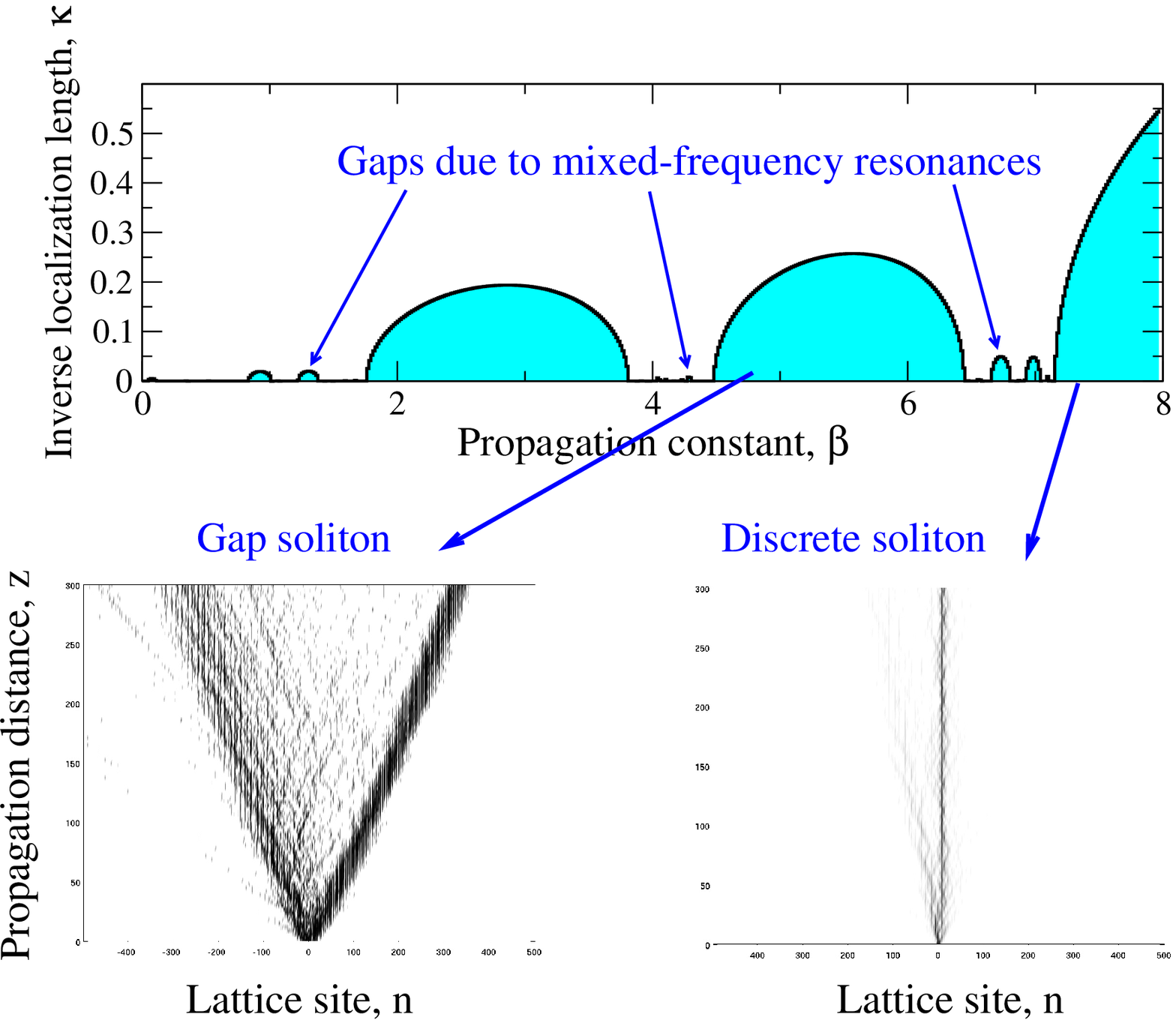}{solitonHarm}{
Band-gap structure and soliton dynamics in a pure ``harmonic'' lattice, which coupling spectrum contains only the 5 dominant peaks shown in Fig.~\rpict{coupling}(b, top), without aperiodic contributions. Input conditions are the same as in Fig.~\rpict{soliton}.
}

We show in Fig.~\rpict{coupling}(a) characteristic quasi-periodic modulation of the inter-site coupling coefficient calculated using Eqs.~\reqt{ALst},\reqt{coupling}. The Fourier spectrum of the coupling coefficient calculated over 1000 lattice sites contains 5 dominant Fourier components [Fig.~\rpict{coupling}(b, top)], which define long-range quasi-periodic modulation. However, the bottom plot in Fig.~\rpict{coupling}(b) clearly demonstrates that the spectrum structure is much more complicated, containing a continuous set of smaller components, which describe weaker {\em aperiodic modulations} of the lattice. 

In order to investigate the effect of deterministic aperiodic modulations, we first analyze propagation of small-amplitude waves in the form $E_n \exp(i \beta z)$, where the spectral parameter $\beta$ has the meaning of the propagation constant of frequency for optical waves, or negative chemical potential for matter waves. For each value of $\beta$, linearized Eq.~\reqt{dnls} admits two linearly-independent solutions, $E_n^+$ and $E_n^-$. Since the linear spectrum is regular being connected to integrable equations, $E_n^\pm$ are either both extended indicating the presence of transmission band, or exponentially localized when $\beta$ is inside a band-gap. In the latter case, we can choose the mode indices such that $\lim_{n\rightarrow \pm\infty} E_n^{\pm} = 0$, and it can be demonstrated using the transfer-matrix approach~\cite{Yeh:1988:OpticalWaves} that their exponential decay rate is the same. Numerically, we calculate the inverse localization length as follows,
$$\kappa(\beta) = \lim_{n \rightarrow \infty} \log\left[ \max_{n/2,n} |E_n| / \max_{1,n/2} |E_n| \right].$$ 
This dependence defines the linear spectrum: $\kappa \rightarrow 0$ for extended waves inside the bands, and $\kappa>0$ correspond to band-gaps, see an example in Fig.~\rpict{soliton}(top). There is a fundamental total-internal-reflection gap, as well as additional gaps corresponding to Bragg-reflection resonances at the dominant modulation frequencies [cf. Fig.~\rpict{coupling}(b,top)]. Each of these gaps can support bright solitons, that may move through the quasi-periodic lattice in a full analogy with periodic structures~\cite{deSterke:1994-203:ProgressOptics, Eggleton:1996-1627:PRL, Fleischer:2003-23902:PRL, Mandelik:2004-93904:PRL, Neshev:2004-83905:PRL, Eiermann:2004-230401:PRL}, see examples in Figs.~\rpict{soliton}(bottom). These simulations are performed for a self-focusing Kerr-type nonlinearity, ${\cal F} = |E_n|^2$, however similar results are expected for other nonlinear dependencies.
We note that some radiation is observed at the initial stage due to non-ideal input conditions modelling  one-beam~\cite{Fleischer:2003-23902:PRL} and two-beam~\cite{Mandelik:2004-93904:PRL, Neshev:2004-83905:PRL} experimental arrangements, however radiation losses of propagating solitons are practically negligible, resembling radiation in periodic lattices~\cite{Yulin:2003-260402:PRL} that remains exponentially small in a broad parameter region.

For comparison, we also consider a `pure' quasi-periodic lattice, which modulation spectrum contains only the five dominant components shown in Fig.~\rpict{coupling}(b, top). Its band-gap spectrum [Fig.~\rpict{solitonHarm}(top)] contains the same principal gaps as observed in Fig.~\rpict{soliton}(top). However, there also appear multiple additional mini-gaps due to mixed-frequency resonances, which are no longer suppressed by aperiodic modulations. We perform numerical simulations using the same input conditions as in Fig.~\rpict{soliton}(bottom), and observe that presence of mini-gaps strongly affects the soliton motion, see Fig.~\rpict{solitonHarm}(bottom). Indeed, the motion of discrete soliton is arrested as it becomes trapped, similar to effects in disordered systems~\cite{Kartashov:physics/0506167:ARXIV}, and
the gap solitons emit strong radiation during their propagation. 
It is most remarkable that small aperiodic modulations have such a dramatic effect on linear wave propagation and soliton dynamics.

In conclusion, we have demonstrated that linear wave propagation and soliton motion in lattices with exactly quasi-periodic modulation of their parameters can be affected by Bragg resonances at mixed modulation frequencies, leading to soliton trapping or strong radiation losses. However, these negative effects can be suppressed in lattices with additional aperiodic modulations, and we have presented a systematic analytical method for calculating the parameters of such optimized lattices. These results can be used, for example, to design photonic structures, such as waveguide arrays, and arrays of traps for atomic condensates.

A.S. acknowledges useful discussions with Yuri Kivshar and Costas Soukoulis.

\end{sloppy}

\begin{thebibliography}{10}

\bibitem{Joannopoulos:1995:PhotonicCrystals}
J.~D. Joannopoulos {\it et~al.}, {\em Photonic Crystals: Molding the Flow of
  Light} (Princeton University Press, Princeton, 1995).

\bibitem{Russell:1995-585:ConfinedElectrons}
P.~St.~J. Russell {\it et~al.}, ``Photonic Bloch waves and photonic band
  gaps,''  in {\em Confined Electrons and Photons}, E. Burstein and C.
  Weisbuch, eds., (1995), \ pp.\ 585--633.

\bibitem{Cristiani:2002-63612:PRA}
M. Cristiani {\it et~al.}, Phys. Rev. A {\bf 65}, 063612 (2002).

\bibitem{Anker:2005-20403:PRL}
T. Anker {\it et~al.}, Phys. Rev. Lett. {\bf 94}, 020403 (2005).

\bibitem{Zoorob:2000-740:NAT}
M.~E. Zoorob {\it et~al.}, Nature {\bf 404}, 740 (2000).

\bibitem{Sanchez-Palencia:cond-mat/0502529:ARXIV}
L. Sanchez~Palencia and L. Santos, arXiv {\bf \mdseries cond-mat/0502529}
  (2005).

\bibitem{Scarola:cond-mat/0506415:ARXIV}
V.~W. Scarola and S. Das~Sarma, arXiv {\bf \mdseries cond-mat/0506415} (2005).

\bibitem{Notomi:2004-123906:PRL}
M. Notomi {\it et~al.}, Phys. Rev. Lett. {\bf 92}, 123906 (2004).

\bibitem{Hollingworth:2001-36611:PRE}
J.~M. Hollingworth {\it et~al.}, Phys. Rev. E {\bf 64}, 036611 (2001).

\bibitem{Sumetsky:2002-332:OE}
M. Sumetsky {\it et~al.}, Opt. Express {\bf 10}, 332 (2002).

\bibitem{Gredeskul:1992-1:PRP}
S.~A. Gredeskul and Yu.~S. Kivshar, Phys. Rep. {\bf 216}, 1 (1992).

\bibitem{deSterke:1994-203:ProgressOptics}
C.~M. {de Sterke} and J.~E. Sipe, ``Gap solitons,''  in {\em Progress in
  Optics}, E. Wolf, ed., (North-Holland, Amsterdam, 1994), Vol.~XXXIII, \ pp.\
  203--260.

\bibitem{Eggleton:1996-1627:PRL}
B.~J. Eggleton {\it et~al.}, Phys. Rev. Lett. {\bf 76}, 1627 (1996).

\bibitem{Fleischer:2003-23902:PRL}
J.~W. Fleischer {\it et~al.}, Phys. Rev. Lett. {\bf 90}, 023902 (2003).

\bibitem{Mandelik:2004-93904:PRL}
D. Mandelik {\it et~al.}, Phys. Rev. Lett. {\bf 92}, 093904 (2004).

\bibitem{Neshev:2004-83905:PRL}
D. Neshev {\it et~al.}, Phys. Rev. Lett. {\bf 93}, 083905 (2004).

\bibitem{Eiermann:2004-230401:PRL}
B. Eiermann {\it et~al.}, Phys. Rev. Lett. {\bf 92}, 230401 (2004).

\bibitem{Soukoulis:1999-255:WRM}
C.~M. Soukoulis and E.~N. Economou, Waves Random Media {\bf 9}, 255 (1999).

\bibitem{Christodoulides:2003-817:NAT}
D.~N. Christodoulides {\it et~al.}, Nature {\bf 424}, 817 (2003).

\bibitem{Trombettoni:2001-2353:PRL}
A. Trombettoni and A. Smerzi, Phys. Rev. Lett. {\bf 86}, 2353 (2001).

\bibitem{Aubry:1980-133:RAR}
S. Aubry and G. Andre, Ann. Israel Phys. Soc. {\bf 3}, 133 (1980).

\bibitem{Soukoulis:1982-1043:PRL}
C.~M. Soukoulis and E.~N. Economou, Phys. Rev. Lett. {\bf 48}, 1043 (1982).

\bibitem{Sokoloff:1985-189:PRP}
J.~B. Sokoloff, Phys. Rep. {\bf 126}, 189 (1985).

\bibitem{Lindquist:1994-9860:PRB}
B. Lindquist {\it et~al.}, Phys. Rev. B {\bf 50}, 9860 (1994).

\bibitem{Bourgain:2004-75:DCDS}
J. Bourgain, Discret. Contin. Dyn. Syst. {\bf 10}, 75 (2004).

\bibitem{Guenther:cond-mat/0504210:ARXIV}
A. Guenther {\it et~al.}, arXiv {\bf \mdseries cond-mat/0504210} (2005).

\bibitem{Ablowitz:1999-287:PLA}
M.~J. Ablowitz {\it et~al.}, Phys. Lett. A {\bf 253}, 287 (1999).

\bibitem{Suris:1994-281:PLA}
Yu.~B. Suris, Phys. Lett. A {\bf 189}, 281 (1994).

\bibitem{Yeh:1988:OpticalWaves}
P. Yeh, {\em {Optical Waves in Layered Media}} (John Wiley \& Sons, New York,
  1988).

\bibitem{Yulin:2003-260402:PRL}
A.~V. Yulin {\it et~al.}, Phys. Rev. Lett. {\bf 91}, 260402 (2003).

\bibitem{Kartashov:physics/0506167:ARXIV}
Y.~V. Kartashov and V.~A. Vysloukh, arXiv {\bf \mdseries physics/0506167}
  (2005).

\end{thebibliography}
\end{document}